\pdfoutput=1
\documentclass[a4paper,11pt]{article}
\usepackage[utf8]{inputenc}
\usepackage{jcappub}
\usepackage[utf8]{inputenc}
\usepackage{amsmath,amssymb}
\usepackage{graphicx}
\usepackage{multicol}
\usepackage{hyperref}
\usepackage{physics}
\usepackage{lipsum}

\usepackage{comment}

\usepackage{pgfplots}
\usepgfplotslibrary{fillbetween}
\usetikzlibrary{patterns}
\pgfplotsset{compat=1.8}

\newcommand{\be}{\begin{equation}} 
\newcommand{\ee}{\end{equation}}
\newcommand{\bea}{\begin{equation}\begin{aligned}} 
\newcommand{\eea}{\end{aligned}\end{equation}}
\newcommand{\ba}{\begin{eqnarray}}
\newcommand{\ea}{\end{eqnarray}}
\renewcommand{\thetable}{\arabic{table}}

\title{ $\beta$-function reconstruction of \\ Palatini inflationary attractors}

\author[a]{Alexandros Karam,}
\author[b]{Sotirios Karamitsos,}
\author[c]{and Margus Saal}

\emailAdd{alexandros.karam@kbfi.ee}

\emailAdd{sotirios.karamitsos@df.unipi.it} 

\emailAdd{margus.saal@ut.ee}

\vspace{5cm}
\affiliation[a]{Laboratory of High Energy and Computational Physics, 
National Institute of Chemical Physics and Biophysics, R{\"a}vala pst.~10, Tallinn, 10143, Estonia}

\affiliation[b]{Dipartimento di Fisica, Universit\`a di Pisa, Largo Bruno Pontecorvo, 56127, Pisa, Italy}

\affiliation[c]{Laboratory of Theoretical Physics, Institute of Physics, University of Tartu, W. Ostwaldi Str 1, 50411 Tartu, Estonia}

\abstract{
Attractor inflation is a particularly robust framework for developing inflationary models that are insensitive to the details of the potential. Such models are most often considered in the metric formulation of gravity. However, non-minimal models may not necessarily maintain their attractor nature in the Palatini formalism where the connection is independent of the metric. In this work, we employ the $\beta$-function formalism to classify the strong coupling limit of inflationary models in both the metric and the Palatini approaches. Furthermore, we determine the range of values for the non-minimal coupling that lead to theories being observationally indistinguishable in metric and Palatini within current accuracy. Finally, we reconstruct the Jordan frame potential for $\xi$-attractors by imposing an explicit form for the $\beta$-function, demonstrating the effect that the choice of metric or Palatini has on the inflationary observables of the theory.
}
 
\begin{document}

\maketitle

\section{Introduction}

    
Cosmic inflation is one of the most robust paradigms for the description of the very early Universe. A sustained period of exponential expansion resolves the classic puzzles of hot Big Bang cosmology, such as the flatness, horizon, and relic problems \cite{Starobinsky1980, Guth:1980zm, Linde:1981mu}. Most importantly, however, it can also explain the large scale structure we observe today. As the background evolution of the inflaton field $\phi$ drives the accelerated expansion of the Universe, its primordial fluctuations result in inhomogeneities, which in turn result in the nearly scale-invariant spectrum observed in the cosmic microwave background (CMB)~\cite{Salopek1990, Ratra:1987rm, Lyth:1998xn}. Moreover, the spectra of perturbations provide a firm test of the inflationary hypothesis, since they are related by consistency relations reflecting their origin from an inflationary potential, and can be measured in a variety of ways including the analysis of CMB anisotropies~\cite{Ade2016}.
 
The recent Planck survey~\cite{Akrami2018} has severely constrained the values for the inflationary observables, including the scalar spectral index $n_s$ and the tensor-to-scalar ratio $r$. This has ruled out simple inflationary models with monomial potentials, motivating us to consider more sophisticated models. One class of favoured models are non-minimally coupled models featuring terms of the form $\xi \phi^2 R$ (where $R$ is the Ricci scalar), where the non-minimal coupling parameter $\xi$ is a pure number. These models are particularly interesting since non-minimal couplings can be motivated by radiative corrections, even if they are absent at tree level. The same holds for Starobinsky-like terms $\alpha R^2$~\cite{Starobinsky1980}.
  
Inflation was originally formulated in the standard metric approach to gravity, in which the notion of parallel transport (as encoded in the connection $\Gamma^\rho_{\mu\nu}$) is directly linked to the notion of distance (as encoded 
in the metric $g_{\mu\nu}$) through the metric-dependent Christoffel symbols. In Palatini gravity, however, the connection is treated as an independent variable, and the curvature invariants are only expressible in terms of the connection, i.e. $R_{\mu\nu} =  R_{\mu\nu}(\Gamma^\rho_{\mu\nu})$ and  $R  =  R (\Gamma^\rho_{\mu\nu})$, and not in terms of the metric \cite{Palatini1919,Ferraris1982}.  In the context of inflation, the Palatini approach deviates from the standard metric approach by virtue of how theories behave under conformal transformations. This makes the study of Palatini inflation particularly relevant in the case of non-minimally coupled models \cite{Bauer:2008zj, Bauer:2010bu, Tamanini:2010uq, Bauer:2010jg, Rasanen:2017ivk, Tenkanen:2017jih, Racioppi:2017spw, Markkanen:2017tun, Jarv:2017azx, Fu:2017iqg, Racioppi:2018zoy, Carrilho:2018ffi, Kozak:2018vlp, Rasanen:2018fom, Rasanen:2018ihz, Almeida:2018oid, Shimada:2018lnm, Takahashi:2018brt, Jinno:2018jei, Rubio:2019ypq, Bostan:2019uvv, Bostan:2019wsd, Tenkanen:2019xzn, Racioppi:2019jsp, Tenkanen:2020dge, Shaposhnikov:2020fdv, Borowiec:2020lfx, Jarv:2020qqm, Karam:2020rpa, McDonald:2020lpz, Langvik:2020nrs, Shaposhnikov:2020gts, Shaposhnikov:2020frq, Gialamas:2020vto, Verner:2020gfa, Enckell:2020lvn, Reyimuaji:2020goi} (see also~\cite{Olmo:2011uz, Bombacigno:2018tyw, Enckell:2018hmo, Antoniadis:2018ywb, Antoniadis:2018yfq, Tenkanen:2019jiq, Edery:2019txq, Giovannini:2019mgk, Tenkanen:2019wsd, Gialamas:2019nly, Tenkanen:2020cvw, Lloyd-Stubbs:2020pvx, Antoniadis:2020dfq, Ghilencea:2020piz, Das:2020kff, Gialamas:2020snr, Ghilencea:2020rxc, Bekov:2020dww, Dimopoulos:2020pas, Gomez:2020rnq, Karam:2021sno}), since such models can be recast in a minimal form precisely through such transformations.

There has been much recent interest in a specific subset of non-minimal models of inflation, termed \emph{attractor models}, which include the $\alpha$-attractors and the related $\xi$-attractors~\cite{Kallosh:2013hoa}. The latter can be achieved by umposting a non-minimal coupling of the form $f(\phi) \propto 1 + \xi \sqrt{V(\phi)}$ (exemplified by Higgs inflation \cite{Bezrukov2008}) or $f(\phi) \propto \xi \sqrt{V(\phi)}$ (exemplified by induced gravity inflation~\cite{Kaiser1994a}). Attractor models, as the name implies, feature predictions which are insensitive to the shape of the inflationary potential in the strong coupling limit. 
Due to  the presence of non-minimal couplings, these models behave differently in the Palatini formalism~\cite{Jarv:2017azx}.
While we expect them to retain their attractor nature, their Palatini predictions will deviate from their metric ones, possibly ``rescuing'' ruled-out models or leading to the development of novel ones.
 
A particularly useful framework for classifying the so-called ``universality classes'' of inflationary attractors is the $\beta$-function formalism \cite{Binetruy2014, Pieroni2016}. Models belonging to these classes share a common scale invariant limit and are discriminated by the degree to which they deviate from scale invariance. In this respect, this classification of models is reminiscent of the Wilsonian renormalisation group (RG), in the sense that exact de Sitter solutions correspond to fixed points in the RG flow. The $\beta$-function formalism is readily applicable to both minimal and non-minimal theories of inflation, and requires only slight modifications to be effectively used to study how attractor models fare in the Palatini approach. 

The inflationary potential in the Einstein frame (where the non-minimal coupling does not appear in the gravity sector, as opposed to the Jordan frame) is enough to determine all the physical observables predicted by a model~\cite{Martin:2013tda, Jarv:2016sow}. The relation between the inflationary potential and the spectra of scalar and tensor perturbations produced during inflation provides an opportunity for reconstructing the inflaton potential from observations. Consequently, if we are given a particular set of observations (of some accuracy), it is possible to reconstruct the inflaton potential in the Einstein frame.

It is convenient to express the scalar and tensor perturbation spectra in terms of $H(\phi)$ (where $H$ is the Hubble parameter) and its derivatives.  This Hamilton--Jacobi approach to functional reconstruction was developed in \cite{Hodges:1990bf, Copeland:1993jj, Lidsey:1995np}, and it is possible to use it to reconstruct the Jordan frame potential if the form of non-minimal coupling is known. However, the resultant potential will depend on whether the metric or Palatini formalism is adopted. Therefore, reconstructing the potential is a straightforward way to examine how the attractor nature of scalar-tensor theories depends on the formalism under which they are studied.    
 
In this paper, we examine the features of attractor theories of inflation and how these are maintained when switching to the Palatini formalism from the metric formalism and vice versa. We also examine how the $\beta$-function formalism can be used to reconstruct viable inflationary models in the metric and Palatini approaches. In Section~\ref{sec:conftrans}, we present an overview of how the metric and Palatini approach differently affect conformal transformations. In Section~\ref{sec:betaformalism}, we discuss the $\beta$-function formalism and its application to inflation including Palatini gravity. We discuss how the notion of the strong coupling limit is dependent on our choice of formalism in Section~\ref{sec:stronglimit}. We also present the mapping between different Jordan frame potentials that give rise to equivalent predictions in different formulations of gravity. In Section~\ref{sec:convergence}, we examine the domain in which metric and Palatini give \emph{approximately} convergent predictions (within observational bounds), and determine the allowed values of the non-minimal coupling for which this occurs.
In Section~\ref{sec:reconstruction}, we present the reconstruction of both the Einstein frame potential and the Jordan frame potential in metric and Palatini by means of an appropriate choice for the $\beta$-function.
We summarise and discuss our results in Section~\ref{sec:discussion}. 

Throughout this paper, we work in natural units and we use the $(+,\!-,\!-,\!-)$ signature for the metric. Subscripts $1$ and $0$ will denote quantities defined in the metric and Palatini formalisms, respectively.

\section{Conformal transformations in metric and Palatini scalar-tensor theories}
\label{sec:conftrans}

The simplest class of models of inflation features a minimally coupled scalar field with a canonical kinetic term, with an action given by
\begin{align}\ 
S = \int \dd^4 x \sqrt{-g} \left[ - \frac{  R}{2} + \frac{ (\partial \phi)^2}{2} - U  (\phi) \right]\,.
\end{align}
Scalar-tensor theories are an extension of these models, and have been thoroughly studied in the literature, particularly in the context of inflation. They are described by a generic action given as
\begin{align}\label{eq:jordan_action}
S = \int \dd^4 x \sqrt{-g} \left[ - \frac{ f(\phi)R}{2} + \frac{k(\phi)}{2}(\partial \phi)^2 - V  (\phi) \right] \,.
\end{align}
When the non-minimal coupling $f(\phi)$ and the non-canonical kinetic term $k(\phi)$ explicitly appear in the gravity sector, the action is said to be in the \emph{Jordan frame}.  In the following, the potential $V(\phi)$ will be referred to as the \emph{Jordan frame potential}.

One of the reasons for the popularity of scalar-tensor models is that they can be quite readily cast in terms of a minimally coupled theory by way of a conformal transformation. The physical meaning of a conformal transformation has been debated extensively, but even though its role beyond the tree level is not yet clear \cite{Steinwachs:2013tr,Kamenshchik:2014waa,Domenech:2015qoa,Falls:2018olk}, it is generally agreed that at the classical level, a conformal transformation captures the same physics in a different manner~\cite{Faraoni:1998qx, Flanagan:2004bz, Jarv:2007iq, Chiba:2013mha, Postma:2014vaa,  Kuusk:2016rso}, a point of view strengthened by the development of frame-invariant~\cite{Jarv:2014hma, Jarv:2015kga} and frame-covariant approaches for scalar-tensor theories~\cite{Jarv:2016sow, Karam:2017zno, Karamitsos:2018lur}. The limit of general relativity of the scalar-tensor theory can be determined and investigated in the Jordan and Einstein frames~\cite{Jarv:2008eb, Kuusk:2008ak, Jarv:2014laa}. 
 
In general, conformal transformations are local, coordinate-dependent transformations of the metric. In the context of inflation, they are taken to be field-dependent transformations of the metric $g_{\mu\nu}$ to a new metric $\tilde  g_{\mu\nu}$:
\begin{align}\label{conformaltrans}
\tilde g_{\mu\nu} = \Omega(\phi)^2 g_{\mu\nu} \,.
\end{align}
Since the conformal factor $\Omega(\phi)$ is only dependent on the field $\phi$ and not the spacetime coordinate $x$, this transformation does not introduce any new dynamical degrees of freedom. As a result, two theories related by a conformal transformation give rise the same observables, at least in the classical level.

Up until now, our discussion is agnostic as to whether we are working in the metric or Palatini formalism. There is however a difference between performing a conformal transformation in the metric and Palatini approach, which fundamentally stems from the transformation rule for the Ricci scalar
\begin{align} 
R = 2g^{\mu\nu} \left({\Gamma^\rho}_{\mu[\nu,\rho]} + {\Gamma^\sigma}_{\mu[\nu}{\Gamma^\rho}_{\rho]\sigma}\right) \,,
\end{align}
where $\Gamma^\rho_{\mu\nu}$ is the Levi--Civita connection in the Jordan frame.
Furthemore, the Ricci scalar $R=g^{\mu\nu}R_{\mu\nu}[\Gamma,\partial\Gamma]$ is a function of the metric tensor $g_{\mu\nu}$ and the connection~$\Gamma^{\rho}_{\mu\nu}$,
where the square brackets indicate antisymmetrisation. The Christoffel symbols transform in the metric formalism, but not in the Palatini formalism. As a result, the Ricci scalar transforms as follows under the conformal transformation \eqref{conformaltrans}:
\begin{align} 
R \to \widetilde R =  \Omega^{-2} ( R - 6 \delta_\Gamma \, \Omega^{-1} \nabla^2 \Omega) \,,
\end{align}
where we use the following notation:
\begin{align}
\delta_\Gamma =  
\begin{cases} 
1 & \quad   \text{  (metric)} 
\\ 
0 & \quad \text{(Palatini)}
\end{cases}
\end{align}
throughout the rest of the paper. The choice of gravity formulation is therefore reflected in the expression of the Einstein frame connection $\widetilde \Gamma^{\rho}_{\mu\nu}$
\begin{align}\label{eq:conn:J}
\widetilde \Gamma^{\rho}_{\mu \nu} = 
  \Gamma^{\rho}_{\mu \nu}
+ 
\left( 1 - \delta_\Gamma \right) \left[\delta^{\rho}_{\mu} \partial_{\nu} \omega(\phi) +
\delta^{\rho}_{\nu} \partial_{\mu} \omega(\phi) - g_{\mu \nu} \partial^{\rho}  \omega(\phi) \right] \,, \quad  \omega\left(\phi\right)=\ln\sqrt{f(\phi)}  \,.
\end{align}
It is possible to eliminate the non-minimal coupling by virtue of a conformal transformation with $\Omega(\phi)^2 = f(\phi)$. In this case, the action becomes
\begin{align}\label{eq:noncanonical_einstein_action}
S = \int \dd^4 x \sqrt{-g} \left[ - \frac{\widetilde R}{2} + F_\Gamma (\phi)  \frac{(\partial \phi)^2}{2} - U(\phi) \right] \,,
\end{align}
where the derivatives are now taken with respect to $\tilde g_{\mu\nu}$ and $\widetilde R$ is defined in terms of $\widetilde \Gamma^{\rho}_{\mu\nu}$. The non-canonical kinetic term and non-canonical potential are given by
\begin{align}\label{normalisation}
 F_\Gamma (\phi) \equiv   \frac{k(\phi)}{f(\phi)} + \frac{3\delta_\Gamma}{2} \frac{f'(\phi)^2}{f(\phi)^2} \,,
\qquad\qquad
U(\phi) = \frac{V(\phi)}{ f(\phi)^2} \,.
\end{align}
Here, the effects of the non-minimal coupling have been absorbed into the kinetic term.

We may further simplify the action \eqref{eq:noncanonical_einstein_action} via a field redefinition. Assuming that we are in a range of $\phi$ in which the kinetic function is one-to-one, we may unambiguously canonically normalise the field $\phi$ by defining a new field $\varphi$ such that
 \begin{align}\label{partial_gamma}
\left( \frac{d\varphi}{d\phi}\right)^2 =   F_\Gamma (\phi) \,.
\end{align}
Then, by denoting the inverse of the solution to \eqref{partial_gamma} by~$\phi_\Gamma (\varphi)$, we may write the action entirely in terms of the canonical field $\varphi$ as
\begin{align}\label{eq:canonical_einstein_action}
S = \int \dd^4 x \sqrt{-g} \left[ - \frac{\widetilde R}{2} +  \frac{(\partial \varphi)^2}{2} - U_\Gamma (\varphi) \right] \,,
\end{align}
where we define
\begin{align}\label{canonical_potential}
U_\Gamma (\varphi) \equiv \frac{V(\phi_\Gamma (\varphi))}{ f(\phi_\Gamma(\varphi))^2 } \,.
\end{align}
This action is now said to be in the \emph{Einstein frame}. We will refer to $U_\Gamma (\varphi)$ both as the \emph{canonical potential} and the \emph{Einstein frame potential}.

As a straightforward extension to the notation \eqref{canonical_potential}, we may use any quantity $X(\phi)$ written in terms of the original field $\phi$ in order to define its canonical counterpart defined in the Einstein frame:
\begin{align}\label{canonical_quantities}
X_\Gamma (\varphi) \equiv X(\phi_\Gamma (\varphi)) \,,
\end{align}
where the subscript denotes whether we use the metric or Palatini normalisation in \eqref{normalisation}.  

All observational features of a theory are encoded in the Einstein frame potential $U_\Gamma(\varphi)$. As a result, the choice of whether to work in the metric or Palatini formalism leads to a different definition of the normalised field due to \eqref{normalisation}, which ultimately leads to distinct theories. In the next section, we will study the $\beta$-function formalism in both the metric formalism and the Palatini formalism, with the ultimate goal of using it in order to reconstruct the inflationary potential.

\section{The inflationary $\beta$-function formalism in the Palatini approach}
\label{sec:betaformalism}

As noted in the Introduction, there is a clear analogy between inflationary models and the Wilsonian RG picture. This is illustrated in Table~\ref{tab:betaRG}. The $\beta$-function formalism proposed in~\cite{Binetruy2014} and further developed in \cite{ Pieroni2016, Pieroni2016a} is the formalisation of this idea, and has recently drawn considerable attention thanks to its intuitive approach to classifying inflationary models \cite{Garriga2016,Kiritsis2017, Cicciarella2017, Binetruy2017, Gao2018, Fei2017, Cicciarella2018, Berera2018, Adam2019, Kiritsis2019, Cicciarella2019, Vernov2019, Mohammadi2020, Adam2020}.

We will eventually employ the $\beta$-function formalism in order to reconstruct the potential. We will deviate from the reconstruction method outlined in \cite{Lidsey:1995np} in the sense that the particle physics sector of the model, i.e. the specific form of the inflaton potential $V(\phi)$, is not considered as the input parameter. Instead, we will use the simplified version of functional reconstruction by implementing the $\beta$-function formalism in the context of Hamilton--Jacobi approach \cite{Kinney:1997ne} in order to derive the potential in both the Einstein and the Jordan frame. 

\begin{table}[ht]
\centering
\begin{tabular}{ll}
RG & inflation \\
\hline
fixed points &  exact de Sitter (scale-invariant solutions)\\
scaling regions &  inflationary epoch  \\
critical exponents &    scaling exponents of observables
\end{tabular}
\caption{The analogy between inflation and the Wilsonian picture.}
\medskip
 
 \label{tab:betaRG}
\end{table}
 
No matter what precise model we choose, the Hamilton--Jacobi formulation of inflation of Salopek and Bond \cite{Salopek1990} is instrumental in understanding the $\beta$-function formalism. In this approach, the Hubble factor is given as a function of $\varphi$, and the usual cosmological equations of motion are recast as
 \begin{align}
\dot \varphi &= -2H'(\varphi)\,,
\\
U_\Gamma(\varphi) &= 3 H(\varphi)^2 - 2 H'(\varphi)^2 \,,
\end{align}
where the prime indicates derivatives with respect to $\varphi$.
It is important to note that this formulation assumes that the scalar field $\varphi$ evolves monotonically with time, and so can indeed act as a ``clock". 
 
In analogy with quantum field theory, we now define the following two ``RG quantities'': the \emph{superpotential}  $W(\varphi)$ and the \emph{beta function} $\beta(\varphi)$:
 \begin{align}
W_\Gamma(\varphi) &\equiv -2H(\varphi) \,,
\\
\beta_\Gamma (\varphi) &\equiv \frac{\dd\varphi}{\dd\ln a} = \frac{\dot \varphi}{H} = - 2 \frac{\dd \ln W_\Gamma}{\dd\varphi}\,.
\label{eq:beta-fun-def}
\end{align}
We will sometimes drop arguments from this point onward to aid readability when it is clear whether a function should be evaluated at the original field~$\phi$ or the canonical field~$\varphi$. Subscripts differentiate between the Einstein frame quantities in the metric and Palatini formalism, as defined in \eqref{canonical_quantities}.

Since the $\beta$-function encodes the extent to which a theory deviates from pure de Sitter, it has to be related to the first Hubble slow-roll parameter $\epsilon_H$:
\begin{equation}
	\epsilon_{H} = -\frac{\dot{H}}{H^2} = 2 \frac{\, \, H'^2}{H^2} = \frac{1}{2} \, \beta_{\Gamma}^2(\varphi) \,.
\end{equation}
We may then express the usual equations of motion in terms of the $\beta$-function. The Hamilton-Jacobi equation can be written as
\begin{equation}
	U_{\Gamma} = 3 H^2 \left( 1 - \frac{1}{6} \, \beta_{\Gamma}^2(\varphi)  \right) \,.
\end{equation}
The acceleration equation can also be written as follows:
\begin{equation}
	\frac{\ddot{a}}{a} = H^2 \left(1 - \frac{1}{2} \, \beta_{\Gamma}^2(\varphi)  \right) \,
\end{equation}
and the equation of state reads as
 \begin{align}
p_{\varphi} = \left( \frac{\beta_{\Gamma}^2 }{3} - 1\right) \rho_{\varphi} \,.
\end{align}
The exact de Sitter solution is given when $\rho_{\varphi} = - p_{\varphi}$, which corresponds to $\beta = 0$. This is the mathematical expression of the statement that fixed points ($\beta = 0$) correspond to pure scale invariance. Note that parametrising $\beta$ is equivalent to fixing the evolution (RG flow) of the system, but since a particular limiting behaviour (in this case de Sitter) can be reached by multiple models, this corresponds to a \emph{universality class}.

We may express $U_\Gamma(\varphi)$ purely in terms of RG quantities:
 \begin{align}\label{eq:w_beta}
U_\Gamma= \frac{3}{4} W_\Gamma^2 \left(1 - \frac{\beta_\Gamma^2}{6} \right) \,.
\end{align}
Close to the stationary point, we find\footnote{We assume that that the Hubble slow-roll parameter is equal to the potential slow-roll parameters to first order, which occurs for quasi-de Sitter solutions.}
\begin{align}\label{eq:sr_beta_def}
U_\Gamma \sim \frac{3}{4} W_\Gamma^2 \,,
\qquad\qquad
\beta_\Gamma \sim - \frac{\dd\ln U_\Gamma}{\dd\varphi}\,.
\end{align}
The latter equation can also be written through the original field as 
\begin{align}
\beta_\Gamma (\varphi) \sim - \left.\frac{\dd\ln V}{\dd\phi}\right\vert_{\phi = \phi_\Gamma(\varphi)} \,.
\end{align}
This notation indicates that all instances of $\phi$ should be replaced by $\phi_\Gamma(\varphi)$ \emph{after} differentiation. The presence of $\sim$ indicates that an equation holds near the stationary point (i.e. for quasi-de Sitter).
 
The exact solution for $\beta$ (away from the stationary point) is simply found by inverting~\eqref{eq:w_beta} and choosing a branch (we may pick the positive one without loss of generality):
 \begin{align}
\beta_\Gamma(\varphi)  =  \sqrt{ 6 \left( 1 - \frac{4}{3} \frac{ U_\Gamma (\varphi) } {W_\Gamma(\varphi)^2}\right)} \,.
\end{align}
The exact expression for $W_\Gamma$ is given by solving
\begin{align}
2U_\Gamma(\varphi) = \frac{3}{2} W_\Gamma (\varphi)^2 - W_\Gamma'(\varphi)^2 \,.
\label{eq:Hamilton-Jacobi}
\end{align}
This procedure allows us to precisely determine the $\beta$-function, from which all observables can be derived for both metric and Palatini. 

Equation~\eqref{eq:beta-fun-def} implies
\begin{align}
    W_\Gamma (\varphi) = W_{\ f} \exp \left[ - \int^{\varphi}_{\varphi_f} \frac{\beta_\Gamma (\hat{\varphi})}{2} \dd \hat{\varphi} \right] = W_{\ f} \exp \left[ - \int^{\varphi}_{\varphi_f} \frac{\beta_\Gamma^2 \left[\hat{\varphi} (N)\right]}{2} \dd N \right] \,,
\end{align}
where we have defined $W_{\ f} = W_\Gamma (\varphi_f)$ and $\varphi_f$ is some constant to be determined later. This solution can be used to reconstruct the Einstein frame potential $U_\Gamma(\varphi)$ regardless of which formalism we use, after which the Jordan frame potential can be determined by the specific form of the canonicalising function $F_\Gamma (\phi)$.
 
Using~\eqref{eq:Hamilton-Jacobi}, we can write the following exact expression for the Einstein frame potential in terms of the $\beta$-function:
\begin{align}
    U_\Gamma = \frac{3 W_{f}^2}{4} \left( 1 - \frac{\beta_\Gamma^2(\varphi)}{6} \right) \exp \left[ - \int^{\varphi}_{\varphi_f} \beta_\Gamma^2 \left[\hat{\varphi} (N)\right] \dd N \right] \,.
\end{align}
Note that we have not assumed quasi-de Sitter in deriving this expression. In quasi-de Sitter, it reduces to
\begin{align}
    U_\Gamma = \frac{3 W^2_{\ f} }{4} \exp \left[ - \int^{\varphi}_{\varphi_f} \beta^2 \left[\hat{\varphi} (N)\right] \dd N \right] \,.
\end{align}
These results will be useful in the following sections where we will reconstruct the potential using an ansatz for the $\beta$-function.
  
\section{Exact convergence of metric and Palatini attractors} 
\label{sec:stronglimit}
 
There are two broad ways in which the attractor behaviour exhibited in metric and Palatini gravity can converge. 
The first is exact, and uses two different (Jordan frame) potentials for the metric and Palatini case. 
The second is approximate: if we start with the same potential, it is possible to have approximately equal observables, 
but only for certain values of the field or for certain values of the non-minimal coupling. We will see later that if inflation takes part mostly in these domains, the attractor profile of the same 
Jordan frame theory in both metric and Palatini gravity can converge approximately. 
For now, we will focus on the exact convergence of attractors both in general and in the the strong limit, and will examine 
the approximate convergence for specific field values in the next section.

\subsection{Equivalence of metric and Palatini attractors
and the strong coupling limit}

Using the $\beta$-function formalism, we turn our attention to theories which exhibit attractor behaviour in the strong limit and how this behaviour is affected by whether we are working in the Palatini or metric formalism. 
For example, motivated by supergravity embeddings~\cite{Binetruy2017}, we consider a theory given by \eqref{eq:jordan_action} with  
\begin{align}\label{eq:sugramodel}
f (\phi) &= 1 + \xi g(\phi) \,,
\\
V (\phi) &= \lambda^2 g(\phi)^2 \,.
\end{align}
The Einstein frame potential in terms of the original field is therefore found to be
\begin{align}
\label{eq:einsteinpot}
U(\phi) = \frac{ \lambda^2 g(\phi)^2}{[1 + \xi g(\phi)]^2} \,.
\end{align}
Therefore, the $\beta$-function near the stationary point then becomes
\begin{align}
\beta_\Gamma(\varphi) &\sim  -\frac{2 g'(\varphi )}{g (\varphi )[1+\xi g (\varphi)]} 
\\
&\simeq - \frac{2}{\xi} \frac{  g'(\varphi )}{   g^2 (\varphi) } \,,
\end{align}
where $\simeq$ denotes the strong coupling limit and we have suppressed the $\Gamma$ index when it is left general for readability purposes.
 
We now return to the canonicalisation equation, which is the origin of the difference between the Palatini and metric formalisms. 
We use two different non-minimal functions: $f_1(\phi) = 1 + \xi g(\phi)$ for the metric case and $f_0(\phi) = 1 + \xi h(\phi)$ for the Palatini case, and
we can write the expression for $F_\Gamma$ in terms of $g(\phi)$ and $h(\phi)$ as 
\begin{align}
F_1 (\phi) &= \frac{1}{1 + \xi  g(\phi ) }  + \frac{3}{2} \frac{\xi ^2 g'(\phi )^2}{ [1 + \xi  g(\phi )]^2 } \,,\\
F_0 (\phi) &= \frac{1}{1 + \xi  h(\phi ) } \,.
\end{align} 

In order for the canonical Einstein frame potential to be exactly the same between metric and Palatini, we must have $U_0(\phi) = U_1(\phi)$ and the canonicalisation functions must also be the same, i.e. $F_1(\phi) = F_0(\phi)$.
To achieve this, we can conclude  that the functions $g(\phi)$ and $h(\phi)$ must satisfy the relationship
\begin{equation} 
h(\phi) = \frac{ g(\phi ) +  \xi  g(\phi )^2  - \frac{3}{2} \xi  g'(\phi )^2 }{1 +  \xi  g(\phi ) + \frac{3}{2} \xi ^2 g'(\phi )^2} \,.
\end{equation} 
The Einstein frame potential $U(\phi)$ (expressed in the terms of the original field) remains the same if Jordan frame potentials are related as
\begin{align}
\label{eq:V0:V1}
V_0(\phi) =  \frac{f_{0}(\phi)^2 }{f_{1}(\phi)^2} \, V_1 (\phi) \,,
\end{align}
and since $F_0(\phi) = F_1(\phi)$, the canonical potentials in metric and Palatini are the same, i.e. $U_1 (\varphi) = U_0(\varphi)$, 
leading to the same observables. Thus, it is always possible (barring discontinuities in the kinetic  function and potential) to exactly map Palatini attractors to metric attractors and vice versa for all values of the field.

We now turn our attention to the equivalence of metric and Palatini attractors in the strong limit. We note that between the metric and the Palatini formalism, the strong coupling limits are also different:
\begin{align}
\label{eq:F1}
F_1 (\phi) &\simeq  
\frac{2 g(\phi) + 3 \xi g'(\phi)^2}{2 \xi g(\phi)^2} \,,\\
F_0 (\phi) &\simeq \frac{1}{  \xi  h(\phi ) } \,.
\label{eq:F2}
\end{align}

As noted above, if we fix the Jordan frame potential $V (\phi)$ in the same form in both the metric and Palatini case, the observational profiles of the resulting theories will be different in general. 
Conversely, if $g(\phi) \equiv h(\phi)$, we must have a different Jordan frame potential for metric and Palatini to converge, otherwise $g(\phi) = \mathrm{constant}$, which corresponds to a minimal theory\footnote{See~\cite{Jarv:2020qqm}, however, for some specific cases where the choices of the model functions $f(\phi)$, $k(\phi)$, $V(\phi)$ can actually lead to the same predictions for $n_s$ and $r$.}. The same holds for the strong limit: in order to obtain undifferentiated observables in the strong limit regardless of metric or Palatini,  the minimal coupling must be transformed accordingly.
  
In order to obtain  convergence of two theories in the strong limit, we impose $F_0(\phi) \simeq F_1(\phi)$ for different non-minimal coupling functions $f_1 = 1 + \xi_0 g(\phi)$ and $f_0 = 1 + \xi_1 h(\phi)$. This leads to the relation
\begin{equation}
\label{eq:h:general}
h(\phi) = \frac{\xi_1}{\xi_0}\frac{ 2 g(\phi ) -3 \xi_1 g'(\phi )^2+2 \xi_1 g(\phi )^2  }{ 2+2 \xi_1 g(\phi )+ 3 \xi_1^2 g'(\phi )^2 }\,.
\end{equation}
For the strong limit to be congruent between the different formalisms, we must have $\xi_0 = \xi_1$. If we fix $g(\phi)$ and look for the corresponding $h(\phi)$, we have the algebraic relation
\begin{equation}
\label{eq:h:algebraic}
h(\phi) =  \frac{ 2 g(\phi ) -3 \xi  g'(\phi )^2+2 \xi  g(\phi )^2  }{ 2+2 \xi  g(\phi )+ 3 \xi ^2 g'(\phi )^2  } \,.
\end{equation}
If we first fix $h(\phi)$, we instead have a differential equation for $g(\phi)$:
\begin{equation}
h(\phi ) \simeq \frac{2 g(\phi )^2-3 g'(\phi )^2}{3 \xi  g'(\phi )^2} \,.
\end{equation}
When solved, this equation returns the relation between $g(\phi)$ and $h(\phi)$ which allows the theories to be 
simultaneously in the strong limit as $\xi$ grows.

As an illustration, we present the case in which $h(\phi) = \phi^n$. The solution for $g(\phi)$ is found to be 
\begin{align}
g(\phi) \ = 
\begin{cases}
C \exp\left[\pm \frac{2}{n-2} \sqrt{\frac{2}{3\xi}} \phi^{\frac{2-n}{2}}\right]  & n\ne 2 \,  \\
C \phi^{\pm \sqrt{\frac{2}{3\xi}}} & n= 2  \,.
\end{cases}
\end{align}
In particular, for $n=2$, we find that 
\begin{align}
\frac{f_0(\phi)^2}{f_1(\phi)^2} \approx  C^{-2} \phi^{4\pm 2\sqrt{\frac{2}{3\xi}}} \,,
\end{align}
and the potentials are related through \eqref{eq:V0:V1}. Therefore, as $\xi \phi^2 \gg 1$ in the strong limit, both the metric and Palatini cases give the same canonicalisation and therefore, thanks to the relation between the Jordan frame potentials, the same observables.

\subsection{Palatini and metric attractors for Starobinsky-like potential}

We now consider the case when the canonical potential takes on a generalised Starobinsky-like form, given by
\begin{equation}
\label{strong:coupling:starobinsky}
U_{\Gamma}(\varphi) \propto  \left( 1 - e^{-\sqrt{\frac{2}{3 \alpha}} \,  (\varphi_{ {\Gamma}} - \bar{\varphi}_{ {\Gamma}})} \right)^2 \,.
\end{equation}
We drop the proportionality constant throughout this section and use hatted quantities to denote integration constants.
The canonical scalar field can be expressed as
\begin{equation}
\label{strong:coupling:ksf:2}
\varphi_{\Gamma} - \bar{\varphi}_{\Gamma} = - \sqrt{\frac{3 \alpha}{2}} \, \ln (1 - \sqrt{U}_{\Gamma}) 
= -\sqrt{\frac{3 \alpha}{2}} \, \ln \left(1 - \sqrt{\frac{V_{\Gamma}(\phi)}{f_{\Gamma}(\phi)^2}} \right) \,.
\end{equation}
In order to arrive at a general solution, we must analytically integrate $\frac{\dd \varphi}{\dd\phi}$, something 
which is not possible in the general case.
However, for a simple case where $f_{\Gamma}(\phi) =1 + \xi_{\Gamma} \, \phi^2$, along with the assumption of strong 
coupling $\xi \phi^2 \gg 1$ at the very beginning, we get
$f_{ {\Gamma}}(\phi) \approx  \xi_{ {\Gamma}} \,  \phi^2$. 
Taking $k(\phi) = 1$, we can integrate  (by selecting signs and integration constants appropriately)
\begin{align}
\label{strong:coupling:ksf:abi}
 \varphi_{\Gamma} - \bar{\varphi}_{\Gamma}   \ = 
\begin{cases}
{  \sqrt{\dfrac{1 + 6 \xi_{1}}{\xi_{1}}}} \, \ln \left(\dfrac{{\phi}}{\bar{\phi}} \right)   & \text{(metric)} \,, 
\\
 \dfrac{1}{\sqrt{\xi_{0}}} \, \ln \left(\dfrac{{\phi}}{\bar \phi} \right) & \text{(Palatini)} \,.
\end{cases}
\end{align}
This is a different approach to the previous subsection, in which the non-minimal coupling functions were different but the non-minimal coupling was kept fixed. Here, we instead consider the relation between the metric and Palatini coupling constants that allows us to recover the strong limit.

Comparing the relations \eqref{strong:coupling:ksf:abi} and \eqref{strong:coupling:ksf:2}, we can relate 
parameters $\alpha$, $\xi_{1}$ and $\xi_{0}$ as
\begin{equation}
\xi_{1} = \frac{1}{  \frac{3}{2} \alpha - 6 } \,, \qquad \xi_{0} =  \frac{2}{3 \alpha} \,. 
\end{equation}
This means that in strong coupling regime, $\xi_{\Gamma} \rightarrow  \infty$  is achieved at different values of parameter $\alpha$  in metric and Palatini case. 
In the Palatini case, $\alpha \rightarrow 0$ and therefore the canonical potential $U_{\Gamma}(\varphi) \rightarrow \Lambda$ (fixed point, i.e. exact de Sitter one as seen in Table~\ref{tab:betaRG}). In the metric case, it appears at $\alpha \rightarrow 4$. 

In general, the Jordan frame potentials, which share the same canonical potential \eqref{strong:coupling:starobinsky}  are different in the metric and in the Palatini case.
We will get the canonical potential \eqref{strong:coupling:starobinsky}~if
\begin{equation}
V_{1}(\phi) \propto
    \xi_{1}^2 \left(1 - \left(\dfrac{\bar{\phi}}{\phi} \right)^{\sqrt{\frac{(1 + 6 \xi_{1}) \xi_{0}}{\xi_{1}}}}  \right)^2 \phi^4
\end{equation}
in the metric case and
\begin{equation}
\label{strong:coupling:V0}
V_{0}(\phi) \propto
    \xi_{0}^2 \left(1 - \frac{\bar{\phi}}{\phi}  \right)^2 \phi^4
\end{equation}
in the Palatini case.

In order to achieve precise convergence (same canonicalisation), we must choose that the coupling parameters of 
metric case and Palatini case are related as follows:
\begin{equation}
\label{eq:sc:hi0:hi1}
    \xi_{0} = \frac{\xi_{1}}{1 + 6 \xi_{1}} \,.
\end{equation}
In this case, the Jordan frame potential can be written as
\begin{equation}
    V_{\Gamma}(\phi) \propto  \xi_{\Gamma}^2 \left(1 - \frac{\bar{\phi}}{\phi}  \right)^2 \phi^4 \,.
\end{equation}
and potentials $V_{0}$ and $V_{1}$ satisfy the relation \eqref{eq:V0:V1}.
If we use the equation \eqref{eq:h:general} as a check, we get
\begin{align}
\xi_{0} = \xi_{1} \ \frac{1 - 6\xi_{1} + \xi_{1} \phi^2}{1 + \xi_{1} \phi^2 + 6 \xi_{1}^2 \phi^2}
\simeq  \xi_{1} \frac{1 - \frac{6}{\phi^2}}{1 + 6 \xi_{1}} \,.
\end{align}
If we assume that the strong coupling  occurs at high values of the scalar field, we get the condition \eqref{eq:sc:hi0:hi1}.
\bigskip 

We can then examine the strong coupling limit for this class of models, which occurs when
\begin{equation}
\label{strong:coupling:assumption}
\frac{f'(\phi)^2}{f(\phi)^2} \gg \frac{k(\phi)}{f(\phi)} \,.
\end{equation}
In this case, the second term in the canonicalisation function dominates, and the theory can be said to be ``metric dominated" (since the term that only appears in the metric formalism is much larger than the first term).

Taking $k = 1$, we may integrate directly \eqref{partial_gamma} in the metric case to obtain the canonical scalar field $\varphi$ as \begin{equation}
\label{strong:copling:inv:field}
\varphi_1 - \bar{\varphi}_{1} =  \pm \sqrt{\frac{3}{2}} \, \ln  f_{1}(\phi) \,,
\end{equation}
where we focus on the positive branch.
 
Using the relation \eqref{strong:copling:inv:field} we can express the potential \eqref{strong:coupling:starobinsky} as
\begin{equation} 
\label{sc:starobinsky:1}
U_{1}(\phi(\varphi)) = 
\left(1 - f_{1}(\phi)^{-\sqrt{\frac{1}{\alpha}}} \right)^2  = 
\left(f_{1}(\phi)^{\sqrt{\frac{1}{\alpha}}} - 1\right)^2 \, f_{1}(\phi)^{-\frac{2}{\sqrt{\alpha}}} \,,
\\
\end{equation} 
and may easily find the Jordan frame potential to be 
\begin{equation}
\label{sc:JF:V1}
V_{1}(\phi)
\left( f_{1}(\phi)^{\sqrt{\frac{1}{\alpha}}}  -  1 \right)^2  f_{1}(\phi)^{2 \left(\frac{\sqrt{\alpha} - 1}{\sqrt{\alpha}} \right)} \,,
\end{equation}
In the case of $\alpha = 1$, this recovers the condition for a single-field Lagrangian to be Starobinsky-like~\cite{Giudice:2014toa}
\begin{equation}
    V_{1}(\phi) = \left(f_{1}(\phi)  - 1 \right)^2  \,.
\end{equation}
If we choose the function $f_{1}(\phi)$ as
\begin{equation} 
    f_{1}(\phi) = \left(\sqrt{\xi_{1}} \phi + \sqrt{1 + \xi_{1} \phi^2} \right)^{-\sqrt{\alpha}} 
    \simeq \left( 2 \sqrt{\xi_{1}} \, \phi \right)^{-\sqrt{\alpha}} \,,
\end{equation}
we can first constrain the coupling constants as $\xi_{1} = \frac{2}{3\alpha}$. 
Using \eqref{sc:JF:V1} we can express the Jordan frame potential in the metric case as follows
\begin{align}
V_{1}(\phi) \simeq 
 \left( 2 \sqrt{\xi_{1}} \, \phi \right)^{2(1 - \sqrt{\alpha})} \,.
\end{align} 
The next step is to find which Jordan frame potential corresponds to the Palatini case. 
However, in the case of Palatini, the strong coupling does not follow from \eqref{strong:coupling:assumption} because one of the terms is not present. 
As an illustrative example, we take again $f_{0}(\phi) =1 + \xi_{0} \phi^2$ and can integrate in Palatini case 
to get
\begin{equation}
    (\varphi_{0}- \bar{\varphi}_{0}) = \pm \frac{1}{\sqrt{\xi_{0}}} \ln \left( \sqrt{\xi_{0}} \phi + \sqrt{1 + \xi_{0} \phi^2} \right)
\end{equation}
and comparing with \eqref{strong:coupling:ksf:2} we get for Jordan frame potential in the strong coupling limit as
\begin{equation}
    V_{0} \simeq \xi_{0}^3 \, \phi^6 \,.
\end{equation}
In this case, $V_0(\phi)$ and $V_1(\phi)$ will give the same observables with the same non-minimal coupling function, in agreement with \eqref{eq:V0:V1}, which relates the metric and Palatini potentials.

\subsection{Non-canonical kinetic term in the Jordan frame ($k(\phi) \ne 1$)}

We now examine the case of a theory with a non-canonical kinetic term. Since the non-minimal coupling does not multiply the second (metric) term in the canonicalising function~$F(\phi)$, appropriately choosing $k(\phi)$ can satisfy 
\begin{equation}
\label{weak:coupling:assumption}
\frac{f'(\phi)^2}{f(\phi)^2} \ll \frac{k(\phi)}{f(\phi)} \,,
\end{equation}
which is the converse to strong coupling regime \eqref{strong:coupling:assumption}. In this regime, the results are insensitive to the choice of metric and Palatini.

As an example, let us consider a theory in the Einstein frame
with a second order pole in the kinetic term, specified by \cite{Terada:2016nqg}
\begin{equation}
f_\Gamma (\phi) = 1 \,, \quad F(\phi) = \frac{3 \alpha}{2} \frac{1}{\phi^2} \,.
\end{equation} 
Switching to the canonical parametrisation by a redefinition of the scalar field yields
\begin{equation}
\varphi = \pm \sqrt{\frac{3 \alpha}{2}} \, \ln \left(\frac{\phi}{\bar \phi }  \right)  \,.
\end{equation}
If the canonical Einstein frame potential $U(\varphi)$ is a generalised Starobinsky potential, we get the following expression for the Jordan frame potential in both metric and Palatini:
\begin{equation}
V_\Gamma (\phi) \propto \left( 1 - \frac{\phi}{\bar \phi }  \right)^2 \,.
\end{equation}
One can interpret the expression for the potential as an expansion around the value $\phi = 0$,
which corresponds to the pole in the kinetic term $k(\phi)$. 
Moreover, as long as inflation takes place around $\phi=0$, the part of invariant potential that affects observables  
is insensitive to the form of $V(\phi)$ far away from $\phi = 0$. 
Therefore, changing the form of $V(\phi)$ by adding higher order terms in the expansion series does not 
significally affect the form of the canonical Einstein potential in the observed region \cite{Galante2015}.

\section{Approximate convergence of metric and Palatini attractors}
 \label{sec:convergence}
 
In the previous section, we studied the convergence between metric and Palatini attractors for different Jordan frame potentials. Imposing an exact convergence gives rise to a mapping between the non-minimal couplings such that the notion of the strong limit is unaffected by the choice of metric or Palatini. However, from the point of view of particle physics, we in principle assume a particular field that drives inflation with some fixed potential and non-minimal coupling (possibly motivated by radiative corrections, SUSY embeddings, UV completion of some EFT...). After the form of the theory is decided in the Jordan frame, the choice of working in metric or Palatini gravity will lead to different observables. As a result, it is important to understand the conditions in which the metric and Palatini attractors can be observationally distinguished for the same starting theory in the Jordan frame, and the conditions in which they give \emph{approximately} the same observables, as opposed to exactly the same observables in the strong coupling limit studied in the previous section.

\subsection{Domain-dependent metric and Palatini convergence}

We now examine a case in which the metric contribution is suppressed for a given domain, resulting in an approximate convergence between the Palatini and metric formalisms. Consider a general attractor theory where $f = 1 + \xi g(\phi)$. 
In order to delineate between the domains in which the metric term (the second term in $F_\Gamma$) dominates, we define the following ratio:
\begin{align}
\omega = \frac{3 }{2} \frac{  \xi ^2 g'(\phi )^2}{  1+\xi  g(\phi )  }\,.
\end{align}
When $\omega\gg 1$ during most of inflation, the metric term (i.e. the second term in $F_\Gamma$) dominates, which means that there will be a phenomenological difference between the two formalisms. On the other hand, when $\omega\ll 1$ during inflation, the metric term is suppressed and the theory is phenomenologically insensitive to the choice of formalism.

Depending on the details of the model, there will be different metric-dominated and metric-suppressed domains on the real line. For a monomial function $\phi^n$ for which $\xi \phi^n \gg 1$, we find that the critical point for which the domains match $\phi_{\rm crit}$ occurs at $\omega = 1$, which is
\begin{align}
\phi_{\rm crit} \approx  \left(\frac{2}{3}\right)^{\frac{1}{n-2}} \left(\frac{1}{   n^2 \xi }\right)^{\frac{1}{n-2}} \,,
\end{align}
when $n\ne 2$. For $n=2$, we find exactly
\begin{align}
\phi_{\rm crit} =  \frac{1}{\sqrt{6    \xi ^2-\xi }} \,.
\end{align}
Therefore, when inflation occurs at $\phi \gg \phi_{\rm crit}$, the theory is metric-dominated. 
 
We will focus on the $n = 2$ case for now. The end of inflation occurs at
\begin{align}
\phi_{e} = \frac{1}{\sqrt{2}}\sqrt{\frac{\sqrt{192 \delta_\Gamma  \xi ^2+32 \xi +1}}{6 \delta_\Gamma  \xi ^2+\xi }-\frac{1}{6 \delta_\Gamma  \xi ^2+\xi }} \,.
\end{align}
Setting $\phi_{e} = \phi_{\rm crit}$ therefore gives us a lower bound for $\xi$ such that the theory is metric-dominated throughout inflation. This returns
\begin{align}
\xi = \frac{1}{12} \left(  \sqrt{3(1+\delta_\Gamma)}+2\right)\,.
\end{align}
Therefore, $\xi \approx 0.311$ for Palatini and $\xi \approx 0.371$ for metric. As expected, for theories like Higgs inflation \cite{Bezrukov2008, Lerner2010}
where $\xi = {\cal O}(10^4)$ in metric and and $\xi = {\cal O}(10^9)$ in Palatini \cite{Rubio:2019ypq,Shaposhnikov:2020fdv}, there is no convergence between Palatini and metric. However, deviating from the quartic coupling $\lambda \sim 0.1$ can drive $\xi$ down to an extent such that which inflation (up until $N= 60$) occurs below $\phi_{\rm crit}$. We can see this by writing the number of $e$-folds in terms of the field
\begin{align}
N(\phi) =  \frac{\xi  \phi ^4}{16}+\frac{\phi ^2}{8} - 
\frac{1 + 8 \xi + \sqrt{1 +32 \xi + 192 \delta_\Gamma  \xi ^2}}{1+16 \xi + 96 \delta_\Gamma  \xi ^2+\sqrt{1+32 \xi +  192 \delta_\Gamma  \xi ^2}} \,.
\end{align}
Then, solving for $N(\phi_{\rm crit}) = N_*$ for $\xi$ (where $N_*$ corresponds to horizon exit) gives us the value for which the point between the metric-dominated and metric-suppressed domain is reached just at horizon exit, and therefore the choice of metric or Palatini can affect observables. This is not possible to solve in closed form, but for $N_* = 60$, we can numerically find 
\begin{align}\label{eq:xi_crit_phi4}
\xi_{\rm crit}(60) = 0.154 \,,
\end{align}
a value which is equal to four significant places even for the two different formalisms. Therefore, a value of $\xi$ of order $10^{-2}$ or smaller ensures that the kinetic term is dominated by the first term throughout inflation, leading to nearly no differentiation between metric and Palatini. The extent to which this occurs depends on the the model. If we know the form of the Einstein potential, we can compute the difference between the observables calculated in metric and Palatini by looking at the resultant Jordan frame potentials. This allows us to restrict the minimal coupling appropriately such that there are no observable differences between the two. This is the topic of the next subsection.

\subsection{Jordan frame potential in metric and Palatini}

Having examined the conditions necessary for metric and Palatini domination to arise, we can  examine the effect the choice of metric or Palatini has on the Jordan potential. Starting with a Einstein potential derived in the metric formalism, it is possible to find its Palatini counterpart, and vice versa. Moreover, it is possible to identify the domain in parameter space in which these two potentials give convergent predictions within observational bounds.

We begin by assuming that we already know how the canonical potential appears in the metric formalism, i.e. we know the form of  $U_1(\phi)$. We can then write the original Jordan frame potential as
\begin{align}
V(\phi) = f(\phi)^2 U_0 (\varphi_0(\phi))\,.
\end{align}
In terms of the Jordan frame potential, the corresponding expression for $U_1$ is
\begin{align}
U_1 (\varphi) =  \frac{V(\phi_1(\varphi))}{f(\phi_1(\varphi))^2}\,.
\end{align}
Therefore, in terms of $U_0$, we have
\begin{align}
U_1 (\varphi) =  U_0 (\varphi_0\circ \phi_1(\varphi))\,,
\end{align}
where $\circ$ denotes function composition. Similarly, assuming that we have the canonical potential $U_1 (\phi)$ in the metric formalism, the corresponding Palatini potential can be found as
\begin{align}
U_0 (\varphi) =  U_1 (\varphi_1\circ \phi_0(\varphi))\,.
\end{align}
Therefore, knowing $\varphi_\Gamma (\phi)$ along with its inverse $\phi_\Gamma(\varphi)$ allows us to transform between the two potentials. Moreover, for a constant $f(\phi)$, we know that $\varphi_0\circ \phi_1 = \varphi_1\circ \phi_0$ is the identity function, leading to no difference between the two potentials.

Consider once again a theory with non-minimal coupling $f(\phi) = 1+\xi\phi^2$. For a canonicalisation such that $\varphi(\phi = 0)=0$, we find the following expressions, which follow from the approximate normalisations in metric and Palatini for large field values:
\begin{align}
\varphi_0 &=   \frac{1}{\sqrt{\xi}}\ln \phi \,,
\\
\varphi_1 &=    \sqrt{\frac{1+6 \xi  }{\xi   }} \ln\phi \,.
\end{align}
Therefore, we can write explicitly
\begin{align}\label{eq:normalisation_composition}
\varphi_0\circ \phi_1(\varphi) &=    \frac{  \varphi }{ \sqrt{1+6 \xi  }} \,, 
\\
\varphi_1\circ \phi_0(\varphi) &=   \sqrt{1+6 \xi  } \, \varphi \,.
\end{align}
We see that, as expected, a small value of~$\xi$ ensures that the the observables will be unaffected by which formalism we choose. Therefore, for the potential discussed in the previous section, we can arrive at an upper bound for $\xi$ for which it is not possible to observationally distinguish between Palatini and metric slow-roll inflation.

We can apply the above procedure to a generic Starobinsky-like potential which we rewrite more simply as
\be
U(\varphi) \propto  1-\alpha e^{-(\varphi-{\bar \varphi})/\sqrt{\alpha}} \,.
\ee
Switching between metric and Palatini therefore introduces a $\xi$ dependence:
\be 
  U(\varphi) \propto  1-\alpha e^{-\frac{\varphi-{\bar \varphi}}{\sqrt{\alpha(1+6\xi)}}} \,.
\ee
The solution of $N(\varphi)$ substituted back into the transformed potential leads to the following normalisation:
\be 
\frac{U(\varphi)}{\epsilon} \propto \frac{2  N^2  }{\alpha ( 1+ 6   \xi ) } \,,
\ee
where we may approximate $\epsilon \simeq \epsilon_U$ in slow-roll via the potential slow-roll parameters for a canonical potential $U$, defined as
\be  
\epsilon_U = \frac{1}{2} \left( \frac{U'}{U} \right)^2 \,, \qquad \eta_U = \frac{U''}{U} \,.
\ee
Using the relation $\dd N = - H \dd t$, we find the following expression for the potential slow-roll parameter 
\be 
\frac{\dd \ln \epsilon_U}{\dd N} = 2 \eta_U - 4 \epsilon_U \,.
\ee
Then, the scalar spectral index $n_s$ and the tensor-to-scalar ratio $r$ in terms of the slow-roll parameters are given as
\ba 
n_s - 1 &=& 2 \eta_U - 6 \epsilon_U = \frac{\dd \ln \epsilon_U}{\dd N} - 2 \epsilon_U \,, \\
r &=& 16 \epsilon_U \,.
\ea
Therefore, for small values of $\xi$, the ``relative error'' induced by using a different formalism is $\pm 6\xi$ for the spectrum (which is proportional to $U/\epsilon$ and  $\pm 36\xi^2$ for $1-n_s$ and $r$ (due to the two derivatives present in the definition of the Hubble slow-roll parameters). 
 
We can find the relative error of the amplitude of the scalar spectrum and of the spectral tilt $1-n_s$ at a $68\%$ confidence interval through the latest constraints from Planck \cite{Akrami2018}:
\be\label{eq:constraints}
A_s \simeq (2.1 \pm 0.0589) \times 10^{-9} \,, \quad n_s = 0.9649 \pm 0.0042 \,, \quad r < 0.056 \,.
\ee
The minimum value of $\xi$ that leads to an observationally distinct tilt $1 \, \sigma$ is $0.057$, which is larger than the minimum value that comes from the strength of the spectrum. Therefore, the minimum value of~$\xi$ that can be resolved experimentally in the context of metric versus Palatini comes from matching the uncertainty of the strength of the spectrum to $6\xi$, which leads to:
\be \label{eq:xi_min_resolution}
\xi_{\rm min} = 0.00468 \,.
\ee
This value is quite robust: it does not depend on the number of $e$-folds or the value of $\alpha$. Larger values of $\xi$ will lead to predictions that are different enough for the two formalisms that they can be resolved with current observations, with the caveat that the potential is roughly a plateau. This value is smaller than the critical value calculated in \eqref{eq:xi_crit_phi4}: this is to be expected, since the critical value occurs at the interface between the metric-suppressed and metric-dominated domains.

\section{$\beta$-function reconstruction of the potential}
\label{sec:reconstruction}

Reconstructing the inflaton potential is one of the more straightforward ways of probing the phenomenology of inflation \cite{Lidsey:1995np, Gobbetti:2015cya, Koh:2016abf, Yi:2016jqr, Gao:2017uja, Gao:2017owg, Fei:2017fub, Herrera:2020mjh, Barbosa-Cendejas:2017pbo}. We will employ the $\beta$-function formalism to reconstructing the potential by means of imposing an ansatz on the $\beta$-function itself. This will also allow us to explicitly demonstrate the conditions for the convergence of the attractor behaviour in both metric and Palatini models. As a result. we will see that in certain domains, the Palatini and metric potentials approximate each other, giving rise to similar observables. 
 
We proceed to write the observables in terms of the $\beta$-function. We first relate $\varphi$ and the number of $e$-folds $N$ through
\be 
\dd \varphi = \pm \sqrt{2 \epsilon_U} \, \dd N \,.
\label{eq:dphi-dN}
\ee
Then, we can express $\epsilon_U$ in terms of the derivative with respect to the number of $e$-folds as
\be 
\epsilon_U = \frac{1}{2} \frac{\dd  \ln U_\Gamma }{\dd N} \,.
\ee
As a result, the observables become
\ba 
A_s &=& \frac{2 H^2}{\pi^2 r} \simeq \frac{2}{3} \frac{U_\Gamma}{\pi^2 r} \,, \\
n_s - 1 &=& \frac{\dd \ln r}{\dd N} - \frac{r}{8} \,, \\
r &=& 8 \frac{\dd\ln U_\Gamma}{\dd N} = 8 \beta^2 \,. \label{eq:rN}
\ea

We now make the following ansatz for the $\beta$-function: 
\be 
\beta^2(N) = \frac{\alpha}{\left( N + \delta \right)^\gamma}\,,
\ee
where $\gamma > 1$, and $\delta$ accounts for the value $\varphi_e$ of the scalar field at the end of inflation.
Then, the inflationary observables are found to be
\be 
r = \frac{8\alpha}{\left( N + \delta \right)^\gamma} \,,
\qquad
n_s - 1 = - \frac{\gamma}{N + \delta} - \frac{\alpha}{\left( N + \delta \right)^\gamma} \,.
\ee
Using~\eqref{eq:rN}, we obtain the reconstructed potential 
\be \label{eq:rec_potential}
U_\Gamma = M^4 \exp \left[ - \frac{\alpha}{\left( \gamma - 1 \right) \left( N + \delta \right)^{\gamma - 1}} \right] \,,
\ee
where the normalisation scale is given by
\be 
M^4 = \frac{3}{2} \pi^2 A_s r \exp \left[ \frac{\alpha}{\gamma - 1} \left( \frac{r}{8 \alpha}  \right)^{\frac{\gamma - 1}{\gamma}} \right] \,.
\ee
From the end-of-inflation condition $\epsilon_U (N=0) = 1$, we obtain the relation
\be 
\alpha = 2 \delta^\gamma \,.
\ee
Also, from~\eqref{eq:dphi-dN} and~\eqref{eq:rN} we have
\be
    \dd\varphi=\sqrt{\frac{r}{8}}\dd N \,,
\ee
which leads to the following solution for the canonical field in terms of the $e$-folds:
\be\label{eq:rec_inflaton}
\varphi-{\bar \varphi} =\left\{
\begin{aligned}
& \frac{2}{2-\gamma}\sqrt{\alpha}(N+\delta)^{\frac{2-\gamma}{2}} \,  \quad & (\gamma\neq 2) \hphantom{,} \,   \\
&\sqrt{\alpha}\ln(N+\delta) \,  \quad & (\gamma=2) \,,
\end{aligned}
\right.
\ee
where $\bar \varphi$ is an integration constant.
Substituting Eq.~\eqref{eq:rec_inflaton} into Eq.~\eqref{eq:rec_potential}, we get the reconstructed potential
\be
U_\Gamma(\varphi)=\left\{
\begin{aligned}
& M^4 \exp\left[-\lambda\left({\bar \varphi}-\varphi\right)^{\frac{2\gamma-2}{\gamma- 2}}\right]\,  \quad & (\gamma \neq 2) \hphantom{,} \,\\
& M^4\exp\left[-\alpha e^{-(\varphi-{\bar \varphi})/\sqrt{\alpha}}\right]\,  \quad & \   (\gamma=2) \,,
\end{aligned}
\right.
\ee
where
\be
  \lambda=\frac{\alpha}{\gamma-1}\left(\frac{\gamma-2}{2\sqrt{\alpha}}\right)^{\frac{2\gamma-2}{\gamma-2}} \,.
\ee
For the $\alpha$-attractor case $\gamma=2$, the potential reduces to
\be
  U_\Gamma(\varphi)=M^4\exp\left[-\alpha e^{-(\varphi-{\bar \varphi})/\sqrt{\alpha}}\right] \,.
\ee
When $\alpha \ll 1$, this potential reduces to
\be\label{eq:asymptotic_potential}
  U_\Gamma(\varphi)=M^4 \left[1-\alpha e^{-(\varphi-{\bar \varphi})/\sqrt{\alpha}}\right] \,,
\ee
which is the asymptotic behaviour of the T-model and E-model. We then use the Planck constraints from \eqref{eq:constraints}: for $\gamma = 2$ and $N = 60$, we find that the maximal allowed value for~$\alpha$ from the constraint on $r$ is $\alpha_{\rm max} = 28.47$. With these values, we obtain $n_s = 0.9616$ which is well within the bounds. For $\gamma = 2.25$, we find $\alpha_{\rm max} = 84.82$ and $n_s = 0.9585$, while for $\gamma = 1.75$ we find $\alpha_{\rm max} = 9.72$ and $n_s = 0.965$. Values of $\gamma$ which are outside the range $\left[ 1.75, \, 2.25 \right]$ give a scalar spectral index which is excluded at $1 \, \sigma$ by Planck~\cite{Akrami2018}.

\begin{figure}
 \centering
 \includegraphics[width=.72\linewidth]{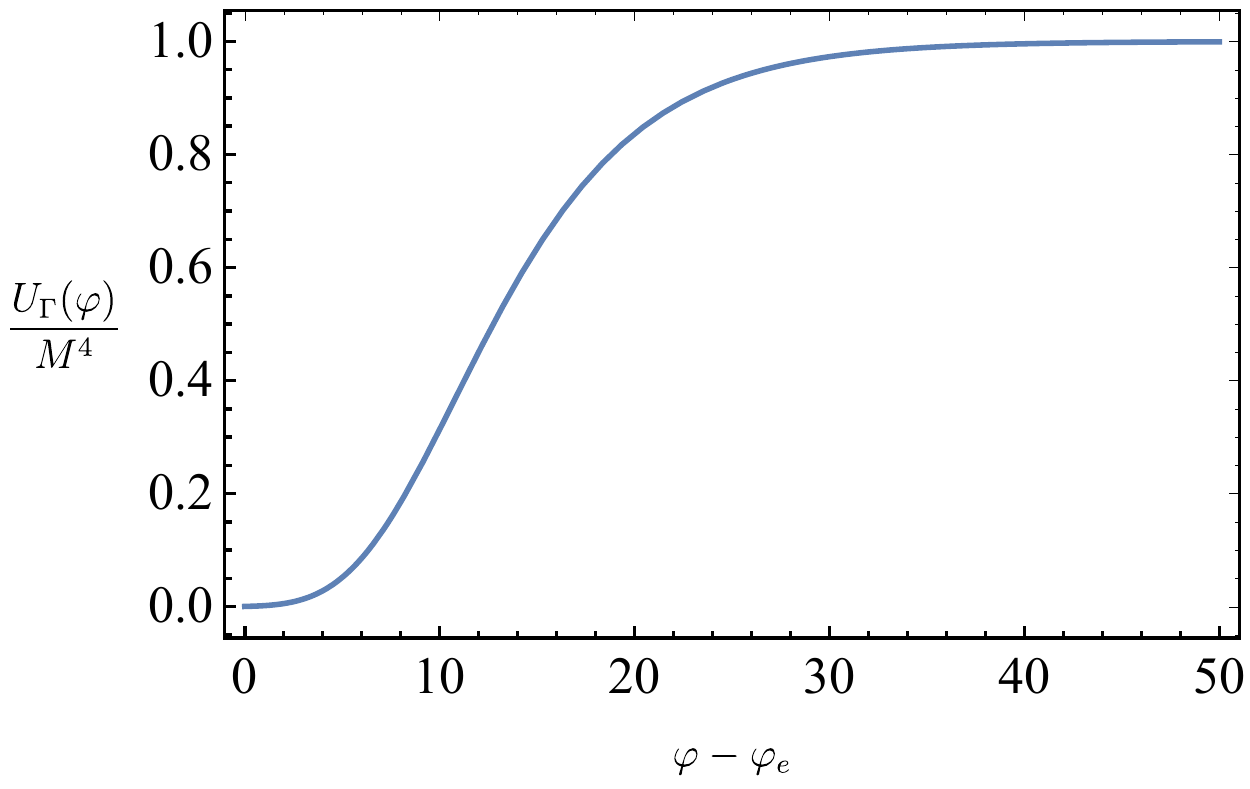}
 \caption{The reconstructed Einstein frame potential for $\gamma = 2$ and $N = 60$. We fixed $\alpha_{\rm max} = 28.47$ such that the upper bound on $r$ from Planck is saturated.}
 \label{fig:potEF}
\end{figure}

In Fig.~\ref{fig:potEF}, we plot the reconstructed Einstein frame potential in the case $\gamma = 2$ and for $N = 60$ and $\alpha_{\rm max} = 28.47$ such that the tensor-to-scalar ratio saturates its maximum allowed value $r_{\rm max} = 0.056$.

\begin{figure}[ht]
\center
\includegraphics[width=.72\textwidth]{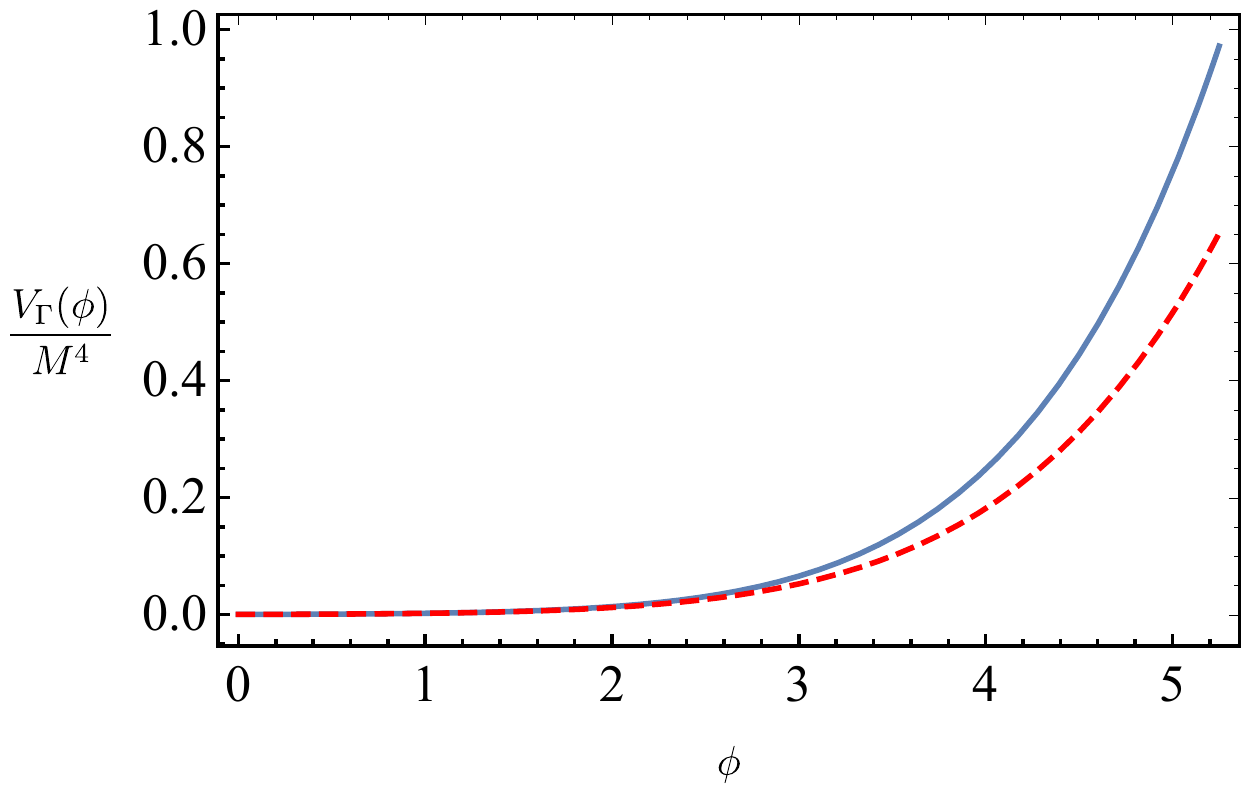}
\caption{The reconstructed Jordan frame potential in metric (solid) and Palatini (red dashed) when 
$\xi = \xi_{\rm crit}$ 
for the case $f(\phi) = 1 + \xi \phi^2$. The potential for $\xi= \xi_{\rm min}$ is not plotted, as it nearly overlaps with the potential in metric gravity, giving rise to nearly identical observables as expected.}
\label{fig:JF-pot}
\end{figure}

Finally, we can relate the reconstructed Einstein frame potential to the Jordan frame potential through
\be 
V_\Gamma(\phi) = f(\phi)^2\, U_\Gamma(\varphi (\phi)) \,,
\ee
If $f(\phi) = 1 + \xi \phi^2$, then the field redefinition gives
\be 
\varphi = \sqrt{\frac{1 + 6 \delta_\Gamma \xi}{\xi}} \sinh^{-1} \left[ \sqrt{\xi} \sqrt{1 + 6 \delta_\Gamma \xi} \phi \right] - \sqrt{6 \delta_\Gamma} \tanh^{-1} \left[ \frac{\sqrt{6 \delta_\Gamma} \xi \phi}{\sqrt{1 + \xi \left( 1 + 6 \delta_\Gamma \xi \right) \phi^2}} \right]\,.
\ee
With this exact equation, it is possible to find the Jordan frame potential in both metric and Palatini in terms of the original field $\phi$. We show the reconstructed metric and Palatini Jordan frame potentials for $\xi_{\rm crit} = 0.154$ (cf.~\eqref{eq:xi_crit_phi4}) in Fig.~\ref{fig:JF-pot}. As expected, the potentials almost entirely overlap for $\xi_{\rm min}$, indicating nearly indistinguishable observables even in the large field limit, whereas for $\xi_{\rm crit}$, they diverge after a certain point (which corresponds to horizon exit).

\section{Summary and discussion}
\label{sec:discussion}

The Palatini formalism of gravitational theories provides an alternative avenue for constructing viable inflationary models. In addition, attractor theories provide some of the most robust models of inflation, and therefore it is interesting to examine how the attractor nature of such models fares in Palatini gravity. Palatini models and metric models of inflation can be related exactly by a change in the Jordan frame potential; however, from a particle physics point of view, the form of the theory in the Jordan frame is set by the UV physics. At that point, making the choice between Palatini or metric gravity will have profound effects on the inflationary phenomenology of the theory.

We have employed the inflationary $\beta$-function formalism in a novel manner 
in order to understand the behaviour of cosmological attractors in both the Palatini and metric formalism. This formalism makes the attractor nature of theories manifest in the same way that the renormalisation group (RG) demonstrates the existence of attractive fixed points. We showed that by making an ansatz for the form of the $\beta$-function, it is possible to recover a class of attractor theories in a straightforward manner.

The strong limit is important in discussions of attractor theories, since it corresponds to the domain in which model-independent results take over and the value of the non-minimal coupling is no longer observationally relevant. We have shown how this limit transforms when changing from metric gravity to Palatini gravity: to maintain the same notion of strong limit, a change of the Jordan frame potential is necessary. If this is not the case, convergence between attractors is approximate, and occurs only for certain values of the field. 

We have further examined the extent to which the reconstruction of the attractor potential is affected by the choice of metric or Palatini. The reconstructed potential from the $\beta$-function is computed in the Einstein frame, and its form in the Jordan frame will depend on both the non-minimal coupling and the chosen formulation of gravity. Therefore, it is possible to place a constraint on the value of the non-minimal coupling for which differentiating between the metric and Palatini formalism is not possible within current observational bounds.

From a formal point of view, the observational effects of choosing either metric and the Palatini are encoded in the Einstein frame potential. 
However, in the context of Palatini versus metric, it becomes necessary to designate either the Jordan or the Einstein frame as more ``fundamental" than the other.
This is because, in the context of particle physics, we expect the Jordan frame to take prominence, as the non-minimal coupling can be motivated by the UV completion of some effective field theory. Therefore, given a fixed (and perhaps unknown) particle theory in which inflation can be realised, it is important to consider a fixed Jordan frame and examine the results for metric or Palatini as they arise in the Einstein frame. Doing so is essentially using cosmology to probe particle physics. On the other hand, the ``fundamental" frame for cosmology is the Einstein frame, where the accelerated expansion is made explicit;
while it is possible to study inflation in the Jordan frame, the equations are considerably more cumbersome. 
Therefore, using cosmology to probe particle physics instead requires us that we go from the Einstein frame to the Jordan frame. As a result, being able to determine which Jordan frame models are compatible with a given Einstein frame model (which can be tested against observations) is important in making the link from cosmology to particle physics.

\acknowledgments
AK was supported by the Estonian Research Council grants MOBJD381 and MOBTT5
and by the EU through the European Regional Development Fund
CoE program TK133 ``The Dark Side of the Universe." SK was supported by ERC grant 669668 NEO-NAT. 
MS was supported by the EU through the European Regional Development Fund
CoE program TK133 ``The Dark Side of the Universe."

\bibliography{References} 

\end{document}